# Typesafe Abstractions for Tensor Operations
# (Short Paper)

Tongfei Chen
Johns Hopkins University, USA
tongfei@cs.jhu.edu

## Abstract
We propose a typesafe abstraction to tensors (i.e. multidimensional arrays) exploiting the type-level programming capabilities of Scala through heterogeneous lists (`HList`), and showcase typesafe abstractions of common tensor operations and various neural layers such as convolution or recurrent neural networks. This abstraction could lay the foundation of future typesafe deep learning frameworks that runs on Scala/JVM.

## 1 Introduction

Recently the machine learning community saw a surge of libraries that handle tensors. Examples include Python libraries such as NumPy [Walt et al. 2011], Theano [Theano Development Team 2016], TensorFlow [Abadi et al. 2015], PyTorch[1], DyNet [Neubig et al. 2017], or Java libraries like Nd4j[2]). These libraries provide abstractions to tensor operations on CPUs or GPUs, and some support automatic differentiation on computational graphs. For tensors in these libraries (belongs to just one type NdArray/Tensor), *specific meaning* is implicitly assigned to each of the dimensions. Correctly manipulating the different dimensions of tensors could be difficult, rendering the whole program prone to runtime errors and hard to reason or maintain: we could mistakenly added up an 1-D tensor and 2-D tensor, or performed matrix multiplication between a 2-D tensor and a 3-D tensor. These errors are only discovered at runtime, or in some libraries, even ignored because of implicit broadcasting. Programmers must keep track of the axes of the tensors themselves (usually as comments in code), rather than leveraging the type system to guide the programmers.

We would like to have a *typing mechanism* for tensors that not only encodes *the rank of the tensor*, but also encodes *what each axis means*. Scala, being a statically-checked type-safe language that is highly capable of type-level programming and DSL construction, as is demonstrated in the popular library Shapeless[3], makes it an ideal language for implementing such typeful and typesafe tensor abstractions.

We describe the design and implementation of such a typesafe abstraction that addresses the typesafety problem of other tensor libraries, and release a prototype.[4]

## 2 Typesafe Tensors

We propose a typesafe tensor abstraction, in which the axes are encoded by a heterogeneous list (`HList`) type parameter.

```scala
trait Tensor[D, A <: HList]
```

`D` is the type of the elements this tensor holds (e.g. `Float`, `Double`, etc.), whereas types in the `HList` type parameter `A` are phantom types, i.e., they only serve as labels.

Basic constructs such as scalars, vectors and matrices could be represented as follows (types `A`, `B` etc. are labels / names to axes).

```scala
type Scalar = Tensor[Float, HNil]
type Vector[A] = Tensor[Float, A :: HNil]
type Matrix[A, B] = Tensor[Float, A :: B :: HNil]
```

Looking at more concrete examples from deep learning applications, images in computer vision, or sentences in which each word is mapped to a word embedding (i.e. vector representation of that word in $\mathbb{R}^d$ space) in natural language processing could be encoded as follows. (See Fig. 1)

```scala
type Image =
  Tensor[Float, Width :: Height :: Channel :: HNil]
type EmbeddedSentence =
  Tensor[Float, Word :: Embedding :: HNil]
```

By encoding the meaning of each axis into the type, our system guarantees that all operations on tensors are allowed only if their operands' axes make sense mathematically. For different mathematical tensor operators, different type guarantees are made. For instance:

- Add (+) guarantees that only two tensors of the exact same axes could be added. Vector[A] and Vector[B] are not addable.
- MatMul (matrix multiplication) guarantees that only two matrices in the shape of Matrix[A, B] and Matrix[B, C] can be multiplied, i.e., the second axis of the first operand and the first axis of the second operand must match.

---

[1] github.com/pytorch/pytorch.
[2] github.com/deeplearning4j/nd4j.

---

[3] github.com/milessabin/shapeless.
[4] github.com/ctongfei/nexus.

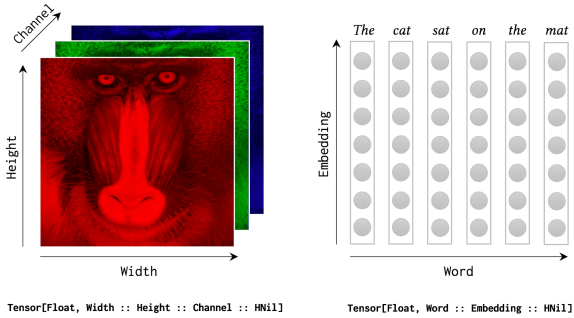

**Figure 1.** Example of encoding an image and a sentence as typesafe tensors.

The general treatment of operator type safety is addressed in Section 5.

Note that the actual size of each axis is not encoded in the type. Shapeless's Church encoding of natural numbers (Nat), when the number is large, significantly reduces compilation speed. Thus, size mismatch (e.g. adding two tensors of the same axes but not the same size) errors will not be captured and will be thrown as runtime errors. We leave the type-level encoding of dimension sizes as future work (possibly by using dependent types).

### 2.1 Shape/axes Manipulation Functions

In common libraries for tensor operations there is usually a set of operations for axes manipulation, namely transpose, expand_dims, squeeze, tile etc. To make these typesafe, we again turn to type-level HList operations.

Take expand_dims as an example. This operation inserts at a specific position a dimension in the tensor with size 1, i.e., a tensor $t$ of shape $(a, b, c)$, when calling expand_dims($t$, axis = 1), results in a tensor of shape $(a, 1, b, c)$. To encode this type relation, we define the following type-level function in the style of Shapeless:

```
trait InsertAt[L <: HList, I <: Nat, X] extends DepFn2[L, X]
{ type Out <: HList }

object InsertAt {
  type Aux[L <: HList, I <: Nat, X, Out0 <: HList] =
    InsertAt[L, I, X] { type Out = Out0 }
```

```
  implicit def at0[T <: HList, H]: Aux[T, _0, H, H :: T] =
    // implementation
  implicit def atN[H, T <: HList, P <: Nat, X, R <: HList]
    (implicit ev: InsertAt.Aux[T, P, X, R]):
    Aux[H :: T, Succ[P], X, H :: R] =
    // implementation
}
```

This type-level function InsertAt[L, I, X] represents the result of the type-level computation that inserts type X to the I-th index of type-level list L. Following the "Aux" pattern, an instance of InsertAt.Aux[L, I, X, R] witnesses that the resulting list is R when inserting X to the I-th index of L.

Given this type encoding, we could define a typesafe method expand_dims for the Tensor[D, A <: HList] trait that admits a new type (label/name for the new axis) and a type-level natural number (Nat) that specifies the position to which the type be inserted. It takes the following type declaration:

```
def expandDims[X, I <: Nat, B <: HList](axis: X, i: I)
  (implicit d: InsertAt.Aux[A, I, X, B], n: ToInt[I]):
  Tensor[D, B] = // implementation
```

This definition of expandDims is completely typesafe: the axes of the output tensor will be completely known at compile time. Other tensor manipulation functions could follow this pattern.

## 3 Computation Graph

For deep learning frameworks, the core algorithm is reverse automatic differentiation [Griewank and Walther 2008; Wengert 1964]. This technique, given the symbolic expression of the loss function of the model, could automatically differentiate through the computation graph and returns the gradient for each parameter in the network.

Computation graphs, being abstract syntax trees (ASTs) of symbolic expressions, are naturally encoded as generalized algebraic data types (GADTs) using case classes [Kennedy and Russo 2005; Xi et al. 2003] in Scala. We propose the following GADT definition:

```
sealed trait Expr[X] { }
case class Input[X]() extends Expr[X]
case class Param[X](var value: X) extends Expr[X]
case class Const[X](val value: X) extends Expr[X]
case class Apply1[X, Y](f: Op1[X, Y], x: X) extends Expr[Y]
case class Apply2[X1, X2, Y](
  f: Op2[X1, X2, Y], x1: X1, x2: X2) extends Expr[Y]
// higher-arities follow
```

In the definition above, Expr[X] is the base trait for all abstract expressions that conceptually hold values of type X.

Input[X] is any input to neural networks that has type X. It is similar to TensorFlow's tf.placeholder, and does not hold any value.

Param[X] represents parameters of neural networks. The values they contain are subject to update during every training iteration of the neural network.

Const[X] represents constants in neural networks. During backpropagation computation of gradients, gradients on Consts are not computed.

Apply1[X, Y](f, x) (or higher-arity generalized Apply2, etc.) nodes are the intermediate result of the application of a differentiable operator $f$ to a symbolic expression $x$. The Op1, Op2 etc. traits are traits for differentiable functions, and will be elaborated in the section below.

## 4   Differentiable Tensor Operators

Reverse automatic differentiation is essentially the application of the chain rule in calculus through the computation graph:

$$\nabla x = \nabla y \cdot \frac{\partial y}{\partial x}.$$

A generic differentiable unary operator $f : X \to Y$ should be capable of performing both forward computation $y = f(x)$ and reverse computation as stated above. It could be encoded as follows (for binary or $n$-ary functions, the definitions can be easily generalized).

```
trait Op1[X, Y] {
  // Performs forward computation y = f(x)
  def forward(x: X): Y
  // Performs backward computation ∇x = ∇y · ∂f/∂x(x)
  def backward(dy: Y, y: Y, x: X): X
}
```

In the backward method, because in machine learning applications the loss function is always a scalar value, the type of $\nabla y$ would be the same as $y$. The reason for including the parameter $y$ in the backward method is because some function's derivative can be easily expressed by the value $y$ rather than $x$. For example, the common sigmoid activation function $y = \sigma(x) = \frac{1}{1+e^{-x}}$ has the derivative $\frac{\partial y}{\partial x} = y(1 - y)$.

## 5   Operator Type Polymorphism

However, an operator can apply to multiple different types. For example, Add(+) can be applied to any two tensors of the same type, however, if their axes do not match, addition is not possible: it makes no sense to add an image of type Tensor[Float, Width :: Height :: HNil] and another transposed Tensor[Float, Height :: Width :: HNil]. Or, matrix multiplication can only apply to any two tensors in which the type of the second axis of the first matrix matches the type of the first axis of the second matrix. Namely, for any types A, B, C, we have matrix multiplication MatMul: (Matrix[A, B], Matrix[B, C]) → Matrix[A, C]. It is obvious that our defined traits OpN are not *polymorphic* to allow this.

To capture arbitrary type relations on inputs/outputs of operators like these above, we define the following polymorphic unary operator, which can be considered as a differentiable version of Shapeless's polymorphic function Poly1, in which the actual implementation of the function is found through implicit resolution of instances of type F[X, Y] (akin to Shapeless's Case.Aux[X, Y]).

```
trait PolyOp1[F[X, Y] <: Op1[X, Y]] {
  // Applies this operator to a symbolic expression
  // requires the actual function F[X, Y] be found
  def apply[X, Y](x: Expr[X])
    (implicit f: F[X, Y]): Expr[Y] = Apply1(f, x)
}
```

This definition captures the following typesafety guarantee: *an operator of type PolyOp1[F] can only be applied to an expression of type Expr[X] if an implicit instance of F[X, Y] is found. If found, the type of the resulting expression is Expr[Y]*. We can arbitrarily define the desired type guarantees in F: for each different tensor operator a different F is defined to express what kinds of operands it can be applied to.

This is easily generalized into higher-arity polymorphic operators. We use two operators to describe the type polymorphism defined above.

### 5.1   Matrix Multiplication

We define a type-polymorphic matrix multiplication operator MatMul:

```
object MatMul extends PolyOp2[MatMulF]
trait MatMulF[X1, X2, Y] extends Op2[X1, X2, Y]
object MatMulF {
  implicit def impl[D, A, B, C]: MatMulF[
      Tensor[D, A :: B :: HNil], Tensor[D, B :: C :: HNil],
      Tensor[D, A :: C :: HNil]] = // implementation
}
```

This essentially expresses: MatMul can only be applied to two expressions Expr[X1] and Expr[X2] only if an instance of MatMulF[X1, X2, Y] is found. This is found only if X1 and X2 take the form of Tensor[D, A :: B :: HNil] and Tensor[D, B :: C :: HNil].

For example, we have three tensors with the following types.

```
val ab: Tensor[Float, A :: B :: HNil]
val ac: Tensor[Float, A :: C :: HNil]
val bc: Tensor[Float, B :: C :: HNil]
```

When compiling MatMul(ab, bc), according to the definition of PolyOp2, the compiler attempts to find the implicit parameter with type MatMulF[Tensor[D, A :: B :: HNil], Tensor[D, B :: C :: HNil], Y]. MatMulF.impl matches this type and by type unification we get type Y being Tensor[D, A :: C :: HNil]. Henceforth the result type of MatMul(ab, bc) is Tensor[D, A :: C :: HNil].

However, when compiling `MatMul(ab, ac)`, the compiler attempts to resolve the implicit parameter with type `MatMulF[Tensor[D, A :: B :: HNil], Tensor[D, A :: C :: HNil], Y]`. Such implicit could not be resolved, resulting in compilation failure, just as we desired. This example shows that using the polymorphic function traits (`PolyOpN`) and implicits, we achieved the axis typesafety for matrix multiplication `MatMul`.

### 5.2 Tensor Contraction

Tensor contraction is also termed variously as `einsum` (Einstein summation) or `tensordot` in various Python libraries. Apart from common usage in deep learning, this operator is also widely found in the sum-product message passing procedure [Pearl 1982] in the belief propagation algorithm for inference in probabilistic graphical models [Koller and Friedman 2009]. Mathematically, given two tensors $A$, $B$, and a list of axis pairs $l$ along which tensors are contracted, we have the result $C$ that retains axes from both $A$ and $B$ not specified in $l$, and all other axes marginalized (summed) out. This general operation subsumes many common linear algebra operations:

- Dot product: $C = \sum_i A_i B_i$;
- Matrix multiplication: $C_{ik} = \sum_j A_{ij} B_{jk}$;
- Tensor product: $C_{i_1 \cdots i_m j_1 \cdots j_n} = A_{i_1 \cdots i_m} B_{j_1, \cdots, j_n}$.

In popular libraries like NumPy or TensorFlow, we write `tensordot(A, B, axes=[[1, 0]])` to specify the axes along which tensors are contracted. Manually writing the axes could be difficult, however using the typeful encoding of tensors, these could be expressed succinctly.

Consider two tensors $A$ and $B$ with $\mathrm{axes}(A) = \{a_1, \cdots, a_m\}$ and $\mathrm{axes}(B) = \{b_1, \cdots, b_n\}$[5]. We define the *natural* tensor contraction $A \bowtie B$ as the contraction of all the axes that share the same name/label (similar to the *natural* join [Codd 1979] in relational databases whereby columns with the same names are joined). For example, the natural tensor contraction of matrices $A[i, j]$ and $B[j, k]$ is the natural matrix multiplication $AB$, but the natural tensor contraction of matrices $A[i, j]$ and $B[k, j]$ is a transposed product $AB^\mathrm{T}$ as the contraction aligns axes with the same name $j$.

What is $\mathrm{axes}(A \bowtie B)$? Because axes that occur in both $A$ and $B$ are contracted, we could see that $\mathrm{axes}(A \bowtie B) = \mathrm{axes}(A) \triangle \mathrm{axes}(B)$, where $\triangle$ is the symmetric set difference. We could define a typelevel function to capture symmetric difference:

```
trait SymDiff[A <: HList, B <: HList] extends DepFn2[A, B]
{ type Out <: HList }
```

Given this, we could encode *natural* tensor contraction typefully and typesafely with `SymDiff` as an implicit evidence as

---
[5]For a tensor $A$ with type `Tensor[D, Axes <: HList]`, denote its axes by $\mathrm{axes}(A) = \mathrm{Axes}$, which is the `HList`.

```
object Contract extends PolyOp2[ContractF]
trait ContractF[X1, X2, Y] extends Op2[X1, X2, Y]
object ContractF {
  implicit def impl[D, A <: HList, B <: HList, C <: HList]
  (implicit C: SymDiff.Aux[A, B, C]):
  ContractF[Tensor[D, A], Tensor[D, B], Tensor[D, C]] =
    // implementation
}
```

Natural tensor contraction also exhibits an elegant property for automatic differentiation (proof omitted):

$$\nabla A = \nabla(A \bowtie B) \bowtie B; \quad \nabla B = \nabla(A \bowtie B) \bowtie A.$$

The equation above typechecks since

$$A = (A \triangle B) \triangle B; \quad B = (A \triangle B) \triangle A.$$

## 6 Common Neural Layers

In deep learning applications, multiples neural layers (function of symbolic expressions) are often stacked together (function composition). There are a collection of common layers that performs certain functions, whose typesafe encodings we describe below.

### 6.1 Fully-connected Layers

One of the most common neural network layer is the fully-connected layer. It is essentially an affine transformation $\mathbf{y} = \mathbf{Wx} + \mathbf{b}$ on the input vector. It could be encoded as

```
case class Affine[D, A, B](
  W: Param[Tensor[D, B :: A :: HNil]],
  b: Param[Tensor[D, B :: HNil]]
) extends
(Expr[Tensor[D, A :: HNil]] => Expr[Tensor[D, B :: HNil]])
```

where the input vector has type `Tensor[D, A :: HNil]` and the output has type `Tensor[D, B :: HNil]`.

### 6.2 Convolutional Layers

Convolutional layers convolves a kernel with the layer input to produce a tensor of outputs. This is widely used in vision/speech/etc. for its shift invariant properties. For the 2-dimensional case common in computer vision, it could be encoded as

```
case class Convolution2D[D, W, H, IC, OC](
  W: Param[Tensor[D, OC :: IC :: HNil]],
  b: Param[Tensor[D, OC :: HNil]]
) extends (Expr[Tensor[D, W :: H :: IC :: HNil]] =>
  Expr[Tensor[D, W :: H :: OC :: HNil]])
```

where the input (an image) has axes width, height and input channel (e.g. RGB), and the output has three axes: width, height and output channel.

### 6.3 Recursive Layers

Sequential recurrent neural networks can be considered as a semiautomaton $(S, I, \tau)$ where $S = \mathbb{R}^h$ is the set of hidden states, $I = \mathbb{R}^d$ is the set of input vectors, and $\tau : (S, I) \to S$ is the transition function, i.e. the recurrent unit. We encode the recurrent unit as a type

```
type RecurrentUnit[D, S, I] =
  (Tensor[D, S :: HNil], Tensor[D, I :: HNil])
  => Tensor[D, S :: HNil]
```

A recurrent unit such as LSTM [Hochreiter and Schmidhuber 1997] would be encoded as a subtype of RecurrentUnit.

Given an input sequence (Seq[Expr[Tensor[D, I :: HNil]]], an example would be a sentence in which each word is represented by a vector representation), we could naturally use Scala's default combinator foldLeft/Right on sequences with a recurrent unit to get the final hidden state (Expr[Tensor[D, H :: HNil]]) and scanLeft/Right to get all the hidden states (Seq[Expr[Tensor[D, H :: HNil]]]). This generalizes to trees (e.g. sentiment analysis on dependency parse trees of sentences [Tai et al. 2015]) where we could define a tree recursion unit and use it with a fold operation (catamorphism [Sheard and Fegaras 1993]) on trees.

## 7 Usability

Exploiting complex libraries such as Shapeless may impose some difficulty to programmers: slow compiling speed (recursive implicit resolution) and confusing compiler error messages.

Since tensors in general machine learning are usually at most 5 dimensions (in video processing, a tensor could have 5 dimensions: batch, time, height, width, color channel), performing type-level HList operations at compile time or runtime are a negligent overhead. Compiling a normal 3-layer neural network just takes an instant.

We have implemented a simple XOR network (2, 2, 2 neurons for each layer) and a simple image classification network for MNIST (784, 300, 100, 10 neurons for each layer). Future work includes deep convolutional networks for image classification; recurrent networks for text annotation; or sequence to sequence transduction tasks such as machine translation. Runtime benchmark would be left as future work, since we do not have a native CPU or GPU underlying implementation as of now.

Compiler errors could be customized by using the Scala @implicitNotFound annotation. Using the matrix multiplication (MatMul) example above, we could annotate as follows.

```
@implicitNotFound("Cannot apply MatMul to ${X1} and ${X2}.")
trait MatMulF[X1, X2, Y] extends Op2[X1, X2, Y]
```

When multiplying two tensors that should not be multiplied (e.g., calling MatMul on two tensors with type Tensor[Float, A::B::HNil] and Tensor[Float, C::B::HNil]), we would get a compiler message "*Cannot apply MatMul to Tensor[Float, ::[A,::[B,HNil]]] and Tensor[Float,* *::[C,::[B,HNil]]]*"[6] located just at the application site of the operator MatMul. Additionally, IDEs (e.g. Scala plugin in IntelliJ IDEA) will also detect this kind of type error while editing. These error reporting mechanisms would greatly aid programmers to identify potential typing errors.

## 8 Related Work

This work is related to the research area of typed linear algebra. Most recent research focused on typing the size of the multidimensional arrays by using type-level integers, or typing the unit of measurements of each dimension, instead of what this work is presenting: typing the meaning of each axis as a label.

[Eaton 2006] implemented a typed linear algebra system in Haskell where dimension sizes are encoded in the type system. [Griffioen 2015] proposed dimensioned matrices/tensors in which each axis is dimensioned with a unit of measurement, and by this the unit of measurements of each dimension of tensors can be determined at compile time. [Muranushi and Eisenberg 2014] extends this idea to be used in astrophysics research and typed the length of each dimension by using type-level integers.

There has been work that implements tensors/neural networks in typesafe languages such as Java (DeepLearning4j[7]) or Haskell (Grenade[8]). None of these achieved the HList-backed typefulness and typesafety as described in this paper.

## 9 Conclusion and Future Work

A Scala-based typesafe abstraction around tensors in neural networks is presented in this paper. We have demonstrated the typesafety and expressiveness of our abstraction through various examples that are common in deep learning tasks.

Type-level encoding of axis size is not explored this paper. Methods using literal types or dependent types in Scala are out of scope of this paper and left as future work.

Future work also includes further implementation this framework: adding optimized CPU/GPU operations, implementing automatic batching, various training methods, etc. We hope that this tool would eventually evolve into a fully-fledged library for typesafe deep learning in Scala.

## Acknowledgments

The author thanks the three anonymous reviewers, the shepherd Oleg Kiselyov, and Matthew Francis-Landau, Yuhuan Jiang and Tim Vieira whose suggestions and advice greatly helped improve and clarify this paper.

---

[6] To render the HList infix type :: in a more human-readable way, one could use the infix functionality in the Splain Scala compiler plugin (https://github.com/tek/splain).
[7] github.com/deeplearning4j/deeplearning4j
[8] github.com/HuwCampbell/grenade